\documentclass[12pt]{article}  

\def\fdir{}

\usepackage[dvips]{graphicx,color}
\usepackage{myagums4}
\usepackage{epsfig}

\let\gtrsim\relax
\let\lesssim\relax
\usepackage{amssymb}
\usepackage{mathpazo}

\usepackage[dvipdfmx,colorlinks]{hyperref}
\hypersetup{
  citecolor=blue, 
  linkcolor=blue,
   urlcolor=blue
}

\usepackage[comma,sort&compress,numbers]{natbib}

\newcommand{\tcb}{\textcolor{blue}}
\newcommand{\meth}{\tcb{Supplementary Materials}}
\newcommand{\info}{\tcb{Supplementary Materials}}
\newcommand{\sm}{\mbox{$\sim$}}
\newcommand{\as}{\mbox{$a^{*}$}}
\newcommand{\ads}{\mbox{$a^{**}$}}

\newcommand{\pft}{\mbox{$P_{43}$}}
\newcommand{\cacar}{\mbox{{CaCO$_3$}}}
\newcommand{\D}{\Delta}
\newcommand{\beqn}{\begin{eqnarray}}
\newcommand{\eeqn}{\end{eqnarray}}
\newcommand{\noi}{\noindent}
\newcommand{\scs}{\scriptsize}
\newcommand{\pmo}{\mbox{$^{-1}$}}
\newcommand{\myurl}{\url{www2.hawaii.edu/~zeebe/Astro.html}}
\newcommand{\npurl}{\url{www.ncdc.noaa.gov/paleo/study/26970}}
\newcommand{\pea}{\mbox{56.01}}
\newcommand{\ped}{\mbox{170}}
\newcommand{\per}{\mbox{$\pm$30}}
\newcommand{\agu}{\mbox{58}}
\newcommand{\edf}{\tcb{Fig.}}
\def\FigFe1262{\tcb{S4}}
\def  \Fig1262{\tcb{S6}}
\def  \SecParm{\tcb{S6}}
\def   \SecUnc{\tcb{S7}}

\renewcommand{\baselinestretch}{\bls} 

\begin{document}

\title{Solar system chaos and the Paleocene-Eocene boundary age 
constrained by geology and astronomy}

\author{Richard E. Zeebe$^{1,*}$ and Lucas J. Lourens$^2$}
\affil{\vspace*{0.5cm}
     $^*$Corresponding Author.\\ {\small
     $^1$School of Ocean and Earth Science and Technology, 
     University of Hawaii at Manoa, 
     1000 Pope Road, MSB 629, Honolulu, HI 96822, USA. 
     zeebe@soest.hawaii.edu      \\[2ex]
     $^2$Department of Earth Sciences, Faculty of Geosciences, 
     Utrecht University, Princetonlaan 8a,
     3584 CB Utrecht, The Netherlands. L.J.Lourens@uu.nl}  \\[2ex]
     {\it Science}, 10.1126/science.aax0612, 2019. \\ 
     }

\newpage
{\bf
Astronomical calculations reveal the solar system's dynamical 
evolution, including its chaoticity, and represent the 
backbone of cyclostratigraphy and astrochronology. 
An absolute, fully calibrated astronomical time scale 
has hitherto been hampered beyond \sm50~Ma, because orbital 
calculations disagree before that age. Here we present
geologic data and a new astronomical solution (ZB18a), showing 
exceptional agreement from \sm58 to 53~Ma. We provide a new 
absolute astrochronology up to \agu~Ma and a new Paleocene-Eocene 
boundary age ($\pea\pm$0.05~Ma). We show that the Paleocene-Eocene 
Thermal Maximum (PETM) onset occurred near a 405-kyr eccentricity
maximum, suggesting an orbital trigger. We 
also provide an independent PETM duration (\ped\per~kyr) from 
onset to recovery inflection. Our astronomical solution requires 
a chaotic resonance transition at \sm50~Ma in the solar system's 
fundamental frequencies.
}

\newpage
Numerical solutions \footnote{\scs Solutions for Earth's 
orbital eccentricity and inclination are available at \myurl\
and \npurl. We provide results from 100-0~Ma but
caution that the interval 100-58~Ma is unconstrained 
due to chaos.}
for the solar system's orbital motion
provide Earth's orbital parameters in the past, widely used 
to date geologic records and investigate Earth's paleoclimate
\citep{poincare1892,milanko41Natb,hays76,varadi03,laskar04NatB,zeebe17aj,
lourens05,westerhold07,mameyers17,likump18,spalding18}.
The solar system's chaoticity imposes an apparently firm 
limit  of \sm50~Ma on identifying a unique orbital solution, 
as small differences in initial conditions/parameters 
cause astronomical solutions to diverge around that
age (Lyapunov time \sm5~Myr, see \info) \citep{morbidelli02,
varadi03,laskar11ast,zeebe17aj}. Recent evidence for a chaotic
resonance transition (change in resonance pattern, see 
below) in the Cretaceous (Libsack record) 
\citep{mameyers17} confirms the solar system's chaoticity
but, unfortunately, does not provide constraints to
identify a unique astronomical solution.
The unconstrained interval between the Libsack record (90-83~Ma)
and 50~Ma is too large a gap, allowing chaos to drive 
the solutions apart (see \info).
Thus, proper geologic data around 60-50~Ma is essential to 
select a specific astronomical solution and, conversely,
the astronomical solution is essential to extend the 
astronomically calibrated time scale beyond 50~Ma.

We analyzed
color reflectance data (\as, red-to-green spectrum)
\citep{lourens05,westerhold07}
at Ocean Drilling Program (ODP) Site 1262 (see \meth); 
\as-1262 for short,
a proxy for changes in lithology \citep{lourens05}.
The related Fe-intensity proxy \citep{westerhold07}
gives nearly identical
results (\edf~\FigFe1262). We focus on the section 
\sm170-110~m (\sm58-53~Ma), which exhibits a remarkable 
expression of eccentricity cycles at Site 1262 
(refs.~\citenum{lourens05,westerhold07,meyers15,
likump18,meyers19}),
less so in the preceding (older) section.
Our focus interval includes the PETM and Eocene 
Thermal Maximum~2 (ETM2), extreme global warming events
(hyperthermals), considered the best paleo-analogs for the
climate response to anthropogenic carbon release \citep{zachos05,
ipcc13ss,zeebe16ngs}. The PETM's trigger mechanism and 
duration remains highly debated \citep{mcinerney11,roehl07,
murphy10}. Thus, in addition to 
geological and astronomical implications, unraveling the 
chronology of events in our studied interval is critical 
for understanding Earth's past and future climate.

We developed a simple floating chronology, attempting 
to use a minimum number of
assumptions (see \meth). We initially employed a uniform 
sedimentation rate throughout the section (except for 
the PETM) and a root mean square deviation (RMSD)
optimization routine to derive ages (for
final age model and difference from previous work,
see \meth). No additional tuning, 
wiggle-matching, or pre-existing age model was applied 
to the data. 
Using our floating chronology, the best fit 
between the filtered and normalized data target \ads\ 
(a-double-star, Fig.~\ref{FigaStar}) and a given astronomical 
solution was obtained through minimizing the RMSD 
between record and solution by 
shifting \ads\ along the time axis
(offset $\tau$) over a time interval of $\pm200$~kyr,
with ETM2 centered around 54~Ma (see \meth). 
Before applying the minimization, both \ads\ and 
the solution were demeaned, linearly detrended, and normalized 
to their respective standard deviation (Fig.~\ref{FigaStar}). 

It turned out that one additional step was necessary for a 
meaningful comparison between \ads\ and astronomical 
solutions. Relative to all solutions tested here, 
\ads\ was consistently offset (shifted towards the 
PETM after optimizing $\tau$) by \sm{one} short eccentricity cycle
for ages either younger (some solutions) or older than the 
PETM (other solutions). The consistent offset relative 
to the PETM suggests that the condensed PETM interval 
in the data record is the culprit, for which we applied 
a correction, also obtained through optimization.
At Site 1262, the PETM is marked by a \sm16~cm clay layer 
($<$1-wt\% \cacar), largely due to dissolution and some
erosion across the interval \citep{zachos05,zeebezachos07}, 
although erosion of Paleocene (pre-PETM) sediment
alone can not account for the offset of \sm{one} short 
eccentricity cycle (see \meth).
Sedimentation rates were hence nonuniform across the PETM
interval \cite{zachos05,westerhold07,likump18} 
and primary lithologic cycles from variations in 
\cacar\ content are not preserved within the clay layer.
Thus, we corrected the condensed interval by
stretching a total of $k$ grid points 
across the PETM by $\D z$ for a total length of 
$\D L = k \D z$ and included $k$ as a second parameter
in our optimization routine (see Fig.~\ref{FigaStar}). 
Essentially, the correction for the reduction/gap 
in carbonate sedimentation 
across the PETM is determined by the entire record, 
except the PETM itself (see \meth). In summary, we minimized the 
RMSD between data target and solution by a 
stretch-shift operation,
i.e., we simultaneously optimized the number of 
stretched PETM grid points ($k$) and the overall time shift 
($\tau$) between floating chronology and solution.

Our new astronomical solution ZB18a (computations build on 
our earlier work \citep{zeebe15apjA,zeebe15apjB,zeebe17aj}, 
see \meth) agrees exceptionally well with
the final \ads\ record (Fig.~\ref{FigaStar}b) and
has the lowest RMSD of all 18 solutions published to 
date that cover the interval (Table~\ref{TabRMSD}). 
The 18 solutions were computed by multiple investigators,
representing different solution classes due to
initial conditions, parameters, etc.
(see Sections~\SecParm, \SecUnc).
Based on ZB18a, we provide a new astronomically calibrated 
age model to \agu~Ma (see Fig.~\ref{FigaStar}b and
\info) and a revised age for the P/E
boundary (PEB) of $\pea \pm 0.05$~Ma (see \meth\ for errors).
Our PEB age differs from previous ages
\citep{westerhold07,jaramillo10L,charles11,westerhold17} 
but is close to 
approximate estimates from 405-kyr cycle counting across 
the Paleocene \citep{hilgen10} (see \info).

Contrary to current thinking \citep{cramer03,westerhold07,
roehl07,meyers15,westerhold17}, the PETM onset therefore
occurred temporally near, not distant, to a 405-kyr maximum 
in Earth's orbital eccentricity (Fig.~\ref{FigaStar},
cf.\ also ref.~\citenum{likump18}).
As for ETM2 and successive early Eocene hyperthermals
\citep{cramer03,lourens05,zachos10epsl},
this suggests an orbital trigger for the PETM,
given theoretical grounding and extensive, robust 
observational evidence for eccentricity controls
on Earth's climate
\citep{milanko41Natb,lourens05,westerhold07,likump18,
mameyers17,meyers15,charles11,hilgen10,cramer03,roehl07,
westerhold17,zachos10epsl,dunkley18,zeebe17pa}.
Note, however, that the onset does not necessarily
coincide with a 100-kyr eccentricity maximum (see below).
Our analysis also provides an independent
PETM main phase duration of \ped\per~kyr
from onset to recovery inflection (for tie points,
see \edf~\Fig1262\ and \meth). This duration might
be an underestimate, given that sedimentation rates increased 
during the PETM recovery (compacting the recovery would require 
additional stretching of the main phase). 
Our duration is significantly longer than 94~kyr 
(ref.~\citenum{roehl07}) but agrees 
with the $^3$He age model at Site 1266
($167\pm^{34}_{24}$~kyr) \citep{murphy10} and is consistent
with $>$8 cycles in Si/Fe ratios at Zumaia 
\citep{dunkley18}, which, we suggest, are full (not half) 
precession cycles.

If high orbital eccentricity ($e$) also contributed to the 
long PETM duration ($e \simeq 0.025-0.044$ during PETM), then 
the potential for prolonged future warming from eccentricity 
is reduced due to its currently low values 
($e \simeq 0.0024-0.0167$ during next 100~kyr).
A similar argument may hold for eccentricity-related
PETM trigger mechanisms. The PETM occurred superimposed 
on a long-term, multimillion year, warming trend 
\citep{lourens05,zachos10epsl}. Our solution ZB18a shows a 
405-kyr eccentricity maximum around the PETM but reduced 100-kyr 
variability (Fig.~\ref{FigaStar}b). Eccentricity 
in ZB18a remained high prior to the PETM for 
one short eccentricity cycle
(Fig.~\ref{FigaStar}b, arrow), suggesting the combination 
of orbital configuration and background warming 
\citep{zachos10epsl,zeebe17pa} forced
the Earth system across a threshold, resulting in 
the PETM. While similar thresholds may exist in the modern
Earth system, the current orbital configuration (lower $e$) 
and background climate (Quaternary/Holocene) are different 
from 56 million years ago. None of the above, however,
will directly mitigate future warming and is therefore 
no reason to downplay anthropogenic carbon emissions
and climate change.

Our astronomical solution ZB18a shows a chaotic 
resonance transition (change in resonance pattern) 
\citep{laskar11} at \sm50~Ma, visualized
by wavelet analysis \citep{ghil02} of the classical 
variables:\\[-8ex]
\beqn
h = e \ \sin \varpi \qquad ; \qquad
p = \sin (I/2) \ \sin \Omega \ ,
\eeqn \\[-7ex]
where $e$, $I$, $\varpi$, and $\Omega$ 
are eccentricity, inclination, longitude of
perihelion, and longitude of ascending node 
of Earth's orbit, respectively (Fig.~\ref{FigWave}). 
The wavelet spectrum highlights several fundamental 
frequencies ($g$'s and $s$'s) of the solar system, 
corresponding to 
eigenmodes. For example, $g_3$ and $g_4$ are loosely
related to the perihelion precession of Earth's and Mars'
orbits ($s_3$ and $s_4$ correspondingly to the nodes).
The $g$'s and $s$'s are constant in quasiperiodic 
systems but vary over time in chaotic systems (see 
\info). Importantly, the period \pft\ associated with $g_4-g_3$ 
switches from \sm1.5~Myr to \sm2.4~Myr in ZB18a at 
\sm50~Ma, characteristic of a resonance transition
(see arrow, Fig.~\ref{FigWave}) \citep{laskar11}.
Remarkably, an independent analysis of the
\as-1262 record recently also reconstructed 
$\pft \simeq 1.5$~Myr \citep{meyers18}
within the interval \sm56-54~Ma. However,
our individual $g$-values from ZB18a are different
from the reconstructed mean values, although within 
their $2\sigma$ error bounds (see \meth).

Notably, parameters required for long-term integrations 
compatible with geologic observations (e.g., ZB18a vs.\ 
\ads, Fig.~\ref{FigaStar}) appear somewhat incompatible with 
our best knowledge of the current solar system.
For instance, ZB18a is part of a solution class
featuring specific combinations of number of asteroids 
and solar quadrupole moment ($J_2$), with $J_2$
values lower than recent evidence suggests (\info). In
addition, the La10c solution \citep{laskar11} with a small RMSD 
(see Table~\ref{TabRMSD}) used the INPOP08 ephemeris, 
considered less accurate than the more recent 
INPOP10 used for La11 \citep{laskar11ast}. 
Yet, La10c fits the geologic data better than La11 
(Table~\ref{TabRMSD} and ref.~(\citenum{westerhold17})).

The resonance transition in ZB18a is an unmistakable 
manifestation of chaos and also key to distinguishing 
between different solutions before \sm50~Ma, e.g.,
using the $g_4-g_3$ term. This term
modulates the amplitude of eccentricity and, e.g., the 
interval between consecutive minima in a 2-Myr 
filter of eccentricity (Fig.~\ref{FigSmin}). Other 
solutions such as La10c \citep{laskar11} also show a 
resonance transition around 50~Ma. However, the pattern for 
ZB18a is different prior to 55~Ma, which is critical for its
better fit with the data record from 58-53~Ma
(smaller RMSD, see Table~\ref{TabRMSD}, 
Fig.~\ref{FigaStar}). For example, $\pft \simeq 2$ 
and \sm1.6~Myr at \sm59 and \sm56~Ma in La10c
but is rather stable at \sm1.5-1.6~Myr across this interval
in ZB18a. Briefly, to explain the geologic 
record, our astronomical solution requires that the solar
system is (a) chaotic and (b) underwent a specific 
resonance transition pattern between \sm60-50 Myr BP.

\renewcommand{\baselinestretch}{0.95}\selectfont

\small

\newpage
\bibliographystyle{unsrt}


\renewcommand{\baselinestretch}{1.0}\selectfont
\vfill
\noindent
{\small
{\bf Acknowledgments.}
{\bf Funding:}
REZ acknowledges support from the US National Science
Foundation (OCE16-58023). This research was supported by 
the Netherlands Organisation for Scientific Research 
(NWO-ALW 865.10.001) and by the Netherlands Earth System 
Science Centre (NESSC 024.002.001) to LJL. We are grateful 
to the International Ocean Discovery Program for drilling 
and exploring ODP Site 1262.
{\bf Author contributions:}
REZ and LJL devised the study and wrote the manuscript. 
LJL was instrumental in guiding the cyclostratigraphic 
analyses and selecting the a*-1262 record. REZ led the 
numerical integrations for the orbital motion of the 
solar system. 
{\bf Competing interests:}
The authors declare no competing interests. 
{\bf Data and materials availability:}
Solutions for Earth's orbital eccentricity and inclination 
are available at \myurl\ and \npurl. We provide results 
from 100-0~Ma but caution that the interval 100-58~Ma is 
unconstrained due to chaos.
}

\vfill
\noi
{\bf The Supplementary Materials include:}\\[1ex]
Materials and Methods \\
Supplementary Text    \\
Figs. S1 to S8        \\
Tables S1 to S3       \\
References 36-60

\renewcommand{\baselinestretch}{1.7}\selectfont

\clearpage

\def\ex{-1ex}
\begin{table}[hhhhhh]
\caption{RMSD$^a$ between \ads\ record and selected
astronomical solutions.$^b$ \label{TabRMSD}}
\vspace*{5mm}
\begin{tabular}{cccccccc}
\tableline\tableline
Solution  & ZB18a$^{c,d}$ & ZB17a  & ZB17b  & La10c  & La10a  & La11   & Va03   \\
\hline  
 RMSD     & 0.6820        & 0.9108 & 1.0358 & 0.7431 & 0.9854 & 1.0009 & 0.9611 \\
\tableline
\end{tabular}
{\scs \\
$^a$Root mean square deviation. \\[\ex]
$^b$Record and solution were demeaned, detrended, and normalized 
to their standard deviation before calculating RMSD. \\[\ex]
$^c$Z and B in ZB18a derive from {\bf Z}eebe-HN{\bf B}ody  
\citep{zeebe17aj}. \\[\ex]
$^d$Lowest RMSD of 18 solutions published to date:
ZB17a-k ($N=11$) \citep{zeebe17aj}, 
La10a-d (4) \citep{laskar11}, La11 (1) \citep{laskar11ast},
La04 (1) \citep{laskar04NatB},  Va03 (1) \citep{varadi03} 
(7/18 listed).
}
\end{table} \noi

\newpage

\begin{figure}[h]
\caption[]{\small
{\bf Data analysis and comparison of color reflectance \as\
to our astronomical solution ZB18a.}
(a) \as\ at ODP Site 1262 (blue-green), interpolated, 
demeaned, detrended record (Norm. \as) including PETM stretch
(light-blue); scaled long/short eccentricity
cycle filter (blue/gray), 
PETM onset (up-triangle), PETM recovery inflection 
(down-triangle), Elmo (square); mcd = meters composite 
depth. As primary \cacar\ variations 
within the PETM interval are not preserved due to dissolution 
and erosion, the interval was cropped. 
(b) Sum of long and short cycle filter outputs in the time 
domain (data target 
\ads, light blue) and normalized eccentricity of Earth's orbit
from our astronomical solution ZB18a (purple). \ads\ and ZB18a 
were demeaned, detrended, and normalized to their respective 
standard deviation before optimization (RMSD minimization
between \ads\ and solution by stretch-shift 
operation, see text). Across the cropped PETM interval,
\ads\ provides no actual information and is omitted.
Up-triangle and 
error bar indicate our new age for the P/E boundary 
(PEB/PETM onset) of $\pea \pm 0.05$~Ma.
Arrow: prolonged high eccentricity period prior to PETM
(see text).
}
\label{FigaStar}
\end{figure}

\begin{figure}[h]
\caption[]{\small
{\bf Wavelet analysis of astronomical solution.}
Wavelet analysis \citep{ghil02} of (a) $h = e \ \sin \varpi$
and (b) $p = \sin (I/2) \sin \Omega$ from our astronomical 
solution ZB18a (see text). $g$'s and $s$'s indicate fundamental 
frequencies of the solar system's eigenmodes
(multiple frequencies are expressed in the spectrum 
of a single planet). For example, $g_3$ and $g_4$ 
are loosely related to the perihelion precession 
of Earth's and Mars' orbits.
The wavelet amplitude (red = peaks, blue = valleys)
in, e.g., the $g_3$ and $g_4$ frequency band is modulated 
by the period $1/(g_4 - g_3) \simeq 2.4$~Myr for ages 
younger \sm45~Ma, where $g_3 \simeq 
1/74.61$~kyr\pmo\ and $g_4 \simeq 1/72.33$~kyr\pmo\
(ref.~\citenum{zeebe17aj}). Correspondingly,
$1/(s_4 - s_3) \simeq 1.2$~Myr.
However, in our solution, the period associated with
$g_4 - g_3$ switches 
from \sm1.5~Myr to \sm2.4~Myr across the resonance
transition around 50~Ma (arrow). The ratio 
$(g_4 - g_3) : (s_4 - s_3) \simeq$ 1:2 after \sm45~Ma
and closer to 1:1 before but appears irregular.
}
\label{FigWave}
\end{figure}

\begin{figure}[h]
\caption[]{\small
{\bf Resonance transition in selected astronomical solutions.}
Interval between consecutive minima ($\D t_{min}$) in 2-Myr 
Gaussian filter ($\pm$60\%) of Earth's orbital eccentricity 
for selected solutions \citep{laskar11,zeebe17aj}. 
The rise of $\D t_{min}$
around 50~Ma in ZB18a and La10c indicate resonance 
transitions. However, note distinct pattern of ZB18a before 55~Ma.
Hence our solution ZB18a (closest agreement with the data record, 
see Fig.~\ref{FigaStar}) requires that the solar system underwent 
a specific chaotic resonance transition pattern between 
\sm60-50 Myr BP.
}
\label{FigSmin}
\end{figure}


\begin{figure}[p]
\figurenum{1}
\def\epsfsize#1#2{0.75#1}
\hspace*{-12mm} 
\vbox{\epsfbox{\fdir 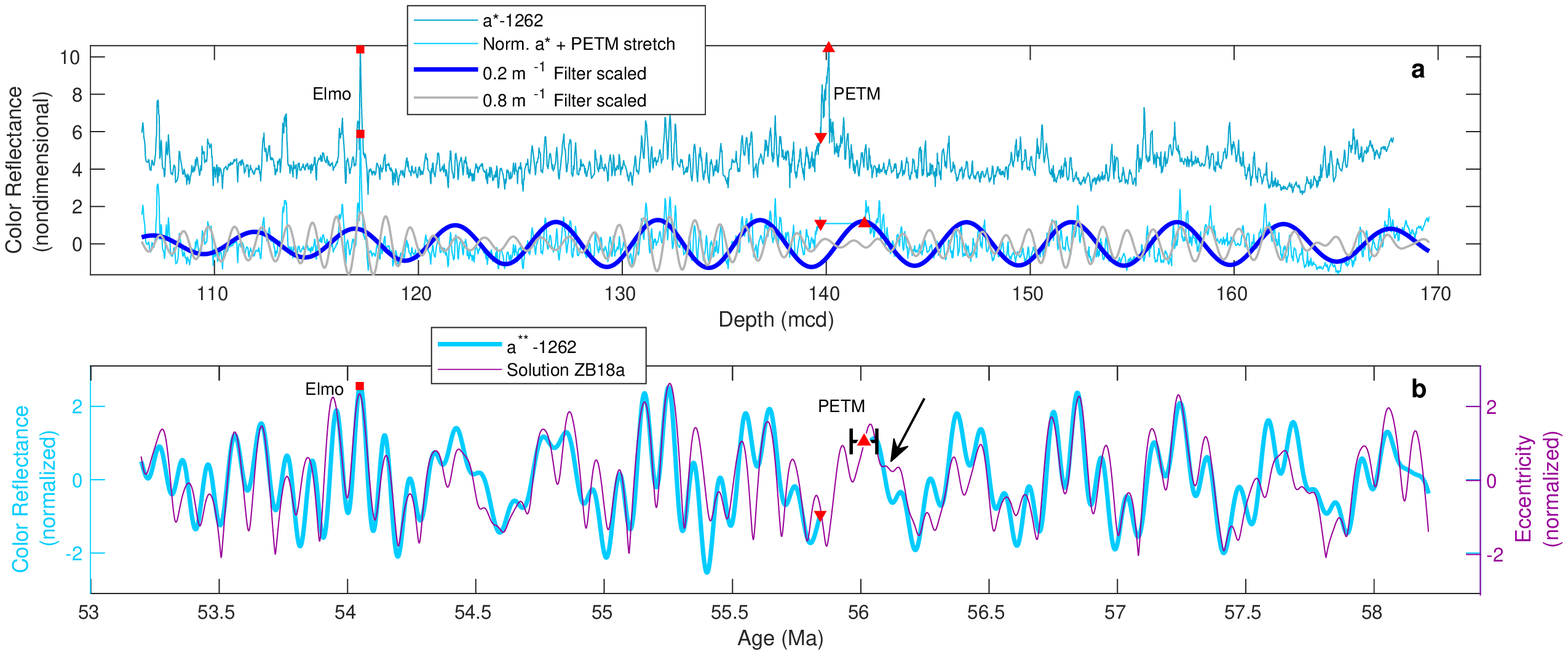}}
\caption[]{}
\end{figure}
\clearpage

\begin{figure}[p]
\figurenum{2}
\def\epsfsize#1#2{0.9#1}
\begin{center}
\hspace*{-2.5cm} 
\vbox{\epsfbox{\fdir 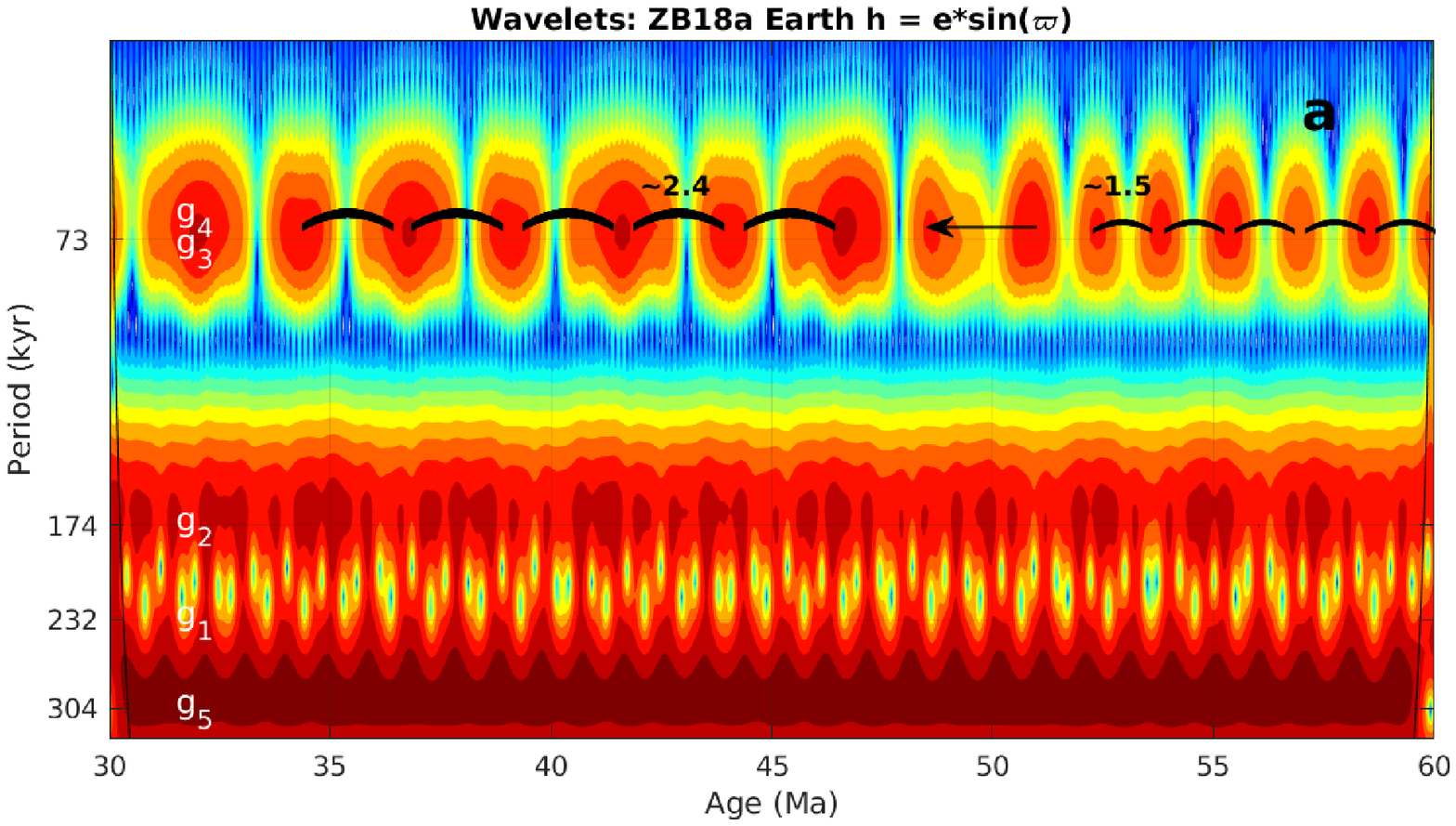}}
\hspace*{-2.5cm} 
\vbox{\epsfbox{\fdir 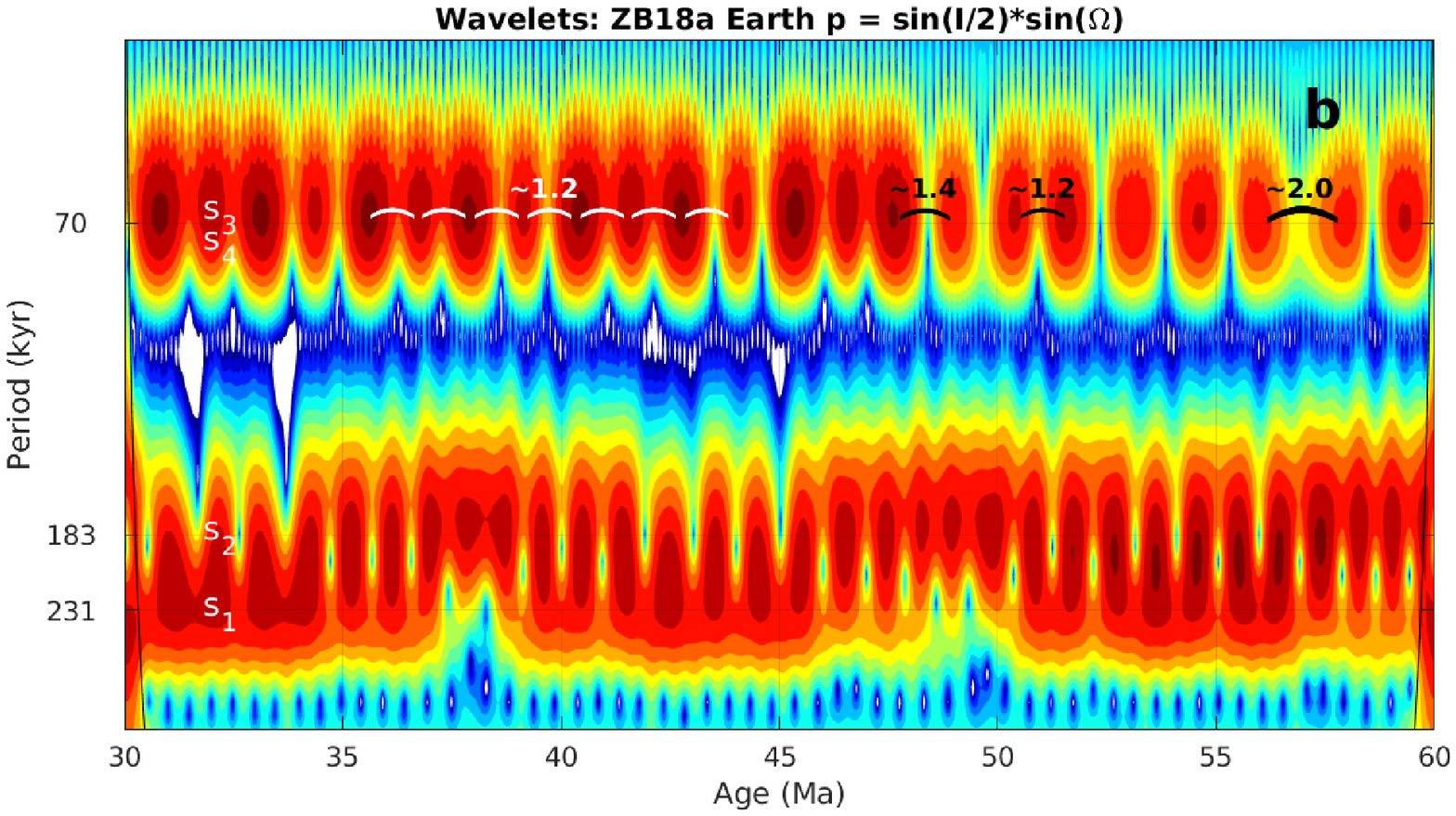}}
\end{center}
\caption[]{}
\end{figure}
\clearpage

\begin{figure}[p]
\figurenum{3}
\def\epsfsize#1#2{1.2#1}
\hspace*{0mm} 
\vbox{\epsfbox{\fdir 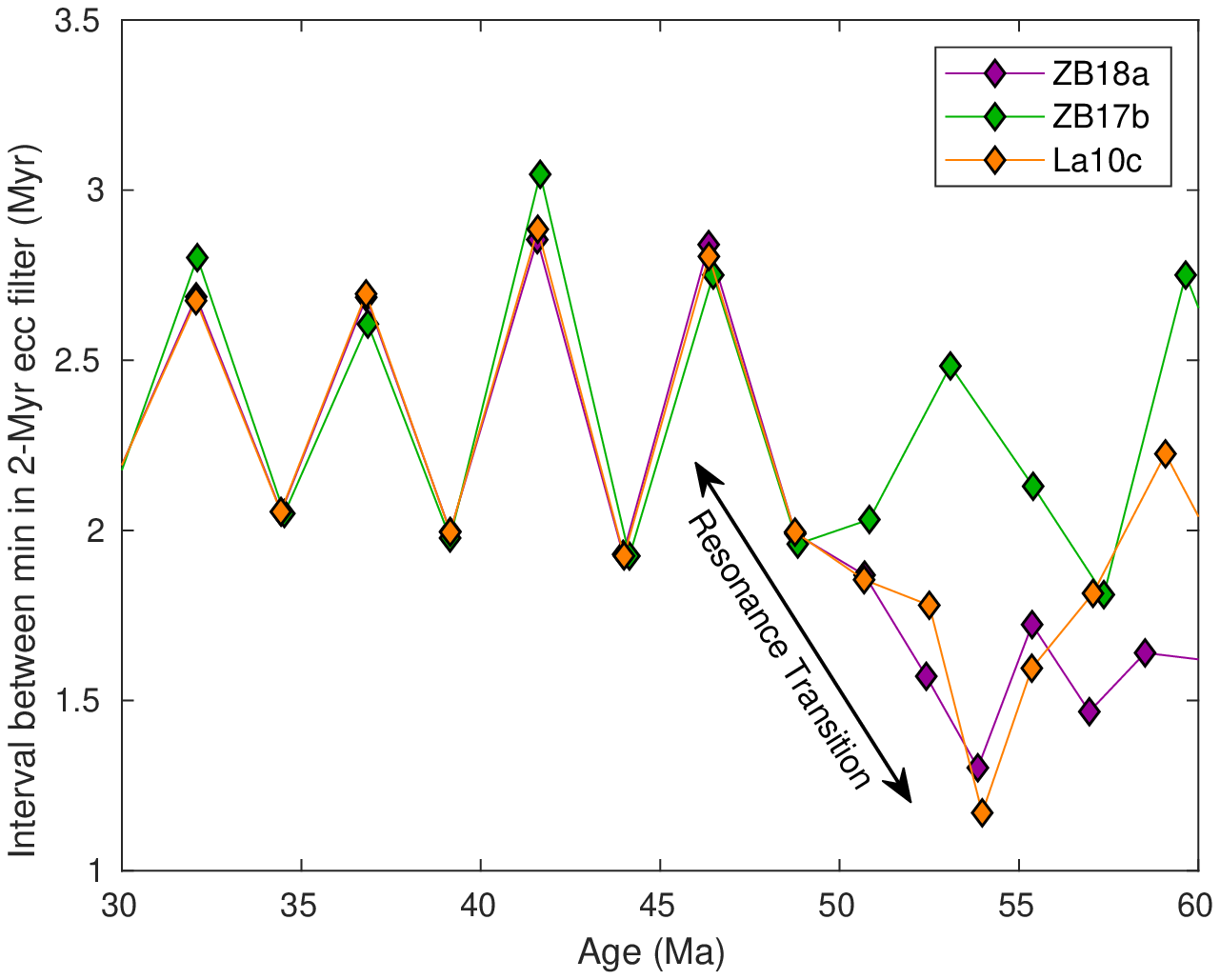}}
\caption[]{}
\end{figure}

\end{document}